\documentclass[reprint,prl,aps,epsfig,longbibliography,superscriptaddress,twocolumn]{revtex4-2}
\usepackage{lineno}
\usepackage{graphicx}
\usepackage[unicode=true,colorlinks=true]{hyperref}
\hypersetup{linkcolor=blue,citecolor=blue,urlcolor=blue}
\usepackage{dcolumn}
\usepackage{bm}
\makeatletter
\usepackage{color, colortbl}
\usepackage{mathtools, nccmath}
\usepackage{soul} 
\definecolor{Gray}{gray}{0.9}

\newcommand{\Rmnum}[1]{\expandafter\@slowromancap\romannumeral #1@}

\begin{document}
\title{Emergent Chiral Metal Phase in Compressible Quantum Hall Fluids}
\begin{abstract}
We report transmitted conductance measurements between a source and reflection-less contacts connected to a compressible quantum Hall fluid with filling fraction $\nu$. We observe that total sum of transmitted conductances universally approaches Hall conductance $\nu(e^2/h)$. The universality of this sum rule is established experimentally across integer and fractional quantum Hall regimes, remaining invariant under variation in temperatures, sample geometries, material qualities and quasi-particle interactions. Chiral transport in compressible quantum Hall fluids, characterized by suppressed dissipative transport with a distinct handedness, is confirmed by floating contact measurements. Consequently, this sum rule emerges as a conduction law of the “chiral metal phase”. Theoretically, we argue that time-reversal-symmetry breaking under a strong magnetic field within this chiral metal gives rise unidirectional trajectories of carriers in a nearly flat potential landscape with point-like disorder potentials, a regime realized when the screening length is smaller than the magnetic length. Within this chiral framework, longitudinal resistance does not originate from bulk dissipation but primarily from the equilibration of electrochemical potentials at the contacts. Our study introduces a new paradigm of chiral transport across a broad class of gapless two-dimensional systems characterized by short-range screening and broken time-reversal symmetry.
\end{abstract}
\author{Suparna Sahoo} 
\affiliation{Saha Institute of Nuclear Physics, 1/AF Bidhannagar, Kolkata 700 064, India}\affiliation{Homi Bhabha National Institute, Anushaktinagar, Mumbai 400094, India}

\author{Suvankar Purkait} 
\affiliation{Saha Institute of Nuclear Physics, 1/AF Bidhannagar, Kolkata 700 064, India}\affiliation{Homi Bhabha National Institute, Anushaktinagar, Mumbai 400094, India}

\author{Pooja Agarwal}
\affiliation{SPEC, CEA, CNRS, Université Paris-Saclay, CEA Saclay, 91191 Gif sur Yvette Cedex, France}
\author{Tanmay Maiti}
\affiliation{Department of Physics and Astronomy, Purdue University, West Lafayette, Indiana 47907, USA}
\author{Sourin Das}
\affiliation{Department of Physical Sciences, IISER Kolkata, Mohanpur, West Bengal 741246, India}
\author{Vladimir Umansky}
\affiliation{Department of Condensed Matter Physics, Weizmann Institute of Science, Rehovot 7610001, Israel}
\author{Biswajit Karmakar}
\email{ biswajit.karmakar@saha.ac.in}
\affiliation{Saha Institute of Nuclear Physics, 1/AF Bidhannagar, Kolkata 700 064, India}
\affiliation{Homi Bhabha National Institute, Anushaktinagar, Mumbai 400094, India}

 \maketitle
\maketitle

A high mobility two-dimensional electron gas (2DEG) subject to a perpendicular magnetic field exhibits integer and fractional quantized Hall conductance plateaus at low temperatures \cite{Klitzing1980_PRL,Tsui1982_PRL}. The integer quantum Hall (IQH) effect arises from quantization of the kinetic energy of electrons \cite{Prange1987}, while the fractional quantum Hall (FQH) effect results from opening of correlated Coulomb gap \cite{Laughlin1983_PRL}, and composite Fermion (CF) picture is used to explain the FQH plateaus \cite{Jain_PhysRevLett.63.199,Jain2007_book}. In the incompressible quantum Hall (QH) plateaus, transmitted conductance is universally measured in terms of quantized values of $(e^2/h)$ as $\nu (e^2/h)$ with filling fraction $\nu = nh/eB$ for carrier density $n$ of the 2DEG under magnetic field $B$, whereas longitudinal conductance $G_{xx}$ becomes zero \cite{Buttiker1986-PRL}. At QH plateaus, integer \cite{Halperin1982_PRB} and fractional \cite{Wen1990_PRB} edge modes carry dissipation-less current. Considering these quantized edge modes, Landauer-Buttiker (L-B) formalism \cite{Buttiker1986-PRL, buttiker1988_backscattering} is widely used to explain the transport properties of 2DEG in complex multi-terminal geometries \cite{sabo2017_nPhy,Maity2020_PRL,Nakamura2023_prl}. 

In the compressible QH region between two QH plateaus, Landau level (LL) is partially filled, where the conducting bulk region is surrounded by the quantized edge modes of the filled LLs \cite{Halperin1982_PRB, Prange1987,Jain2007_book}. The bulk states and the edge modes are generally orthogonal, but in low mobility 2DEG or at high temperatures, they mix by stochastic equilibrium process \cite{Maiti2021_PRB}. In compressible QH regimes, the injected current is distributed among the edge modes and the bulk states. Because of current distribution in the bulk states of compressible QH fluids, standard L-B formalism based on the edge modes is not applicable. In the bulk extended states, quasi-particles suffer elastic or inelastic scattering \cite{buttiker1988_backscattering}, resulting in back-reflection of current and inter-edge backscattering \cite{oswald1998_PhysicaE}. Inelastic scattering such as momentum relaxation \cite{datta_1995} or percolation \cite{trugman1983localization, Percolation_J_T_Chalker_1988} causes dissipation in the bulk, while elastic scattering is associated with non-dissipative transport by energy redistribution \cite{datta_1995}.

The bulk transport is attributed to percolation \cite{trugman1983localization,
Percolation_J_T_Chalker_1988} under disorder screening. However, time reversal symmetry breaking might induce chiral transport in the bulk extended states, but in the presence of disorder, screening, and strong interaction, such chiral transport has never been thoroughly investigated. Since the discovery of the IQH and FQH effects, tremendous effort has been made to understand the transport properties of compressible QH fluids such as the current and voltage distribution \cite{Son1991_PRB, Sirt2025_APL, Sirt2025_NJP, Hirai1994_PRB, Klitzing2001_dist}, plateau to plateau transitions \cite{li2005scaling_PhysRevLett.94.206807, Shahar1997_PRL, wei1994_PRB_scaling}, non-local resistance \cite{Jain1990_NL_PRL, Wang1992_PRB}, edge reconstruction \cite{Purkait2024_PRB1, Purkait2025_PRB2}. 
However, a unified global transport mechanism for compressible QH fluids remains an open fundamental problem.


\begin{figure*}
\includegraphics[width=1 \textwidth]{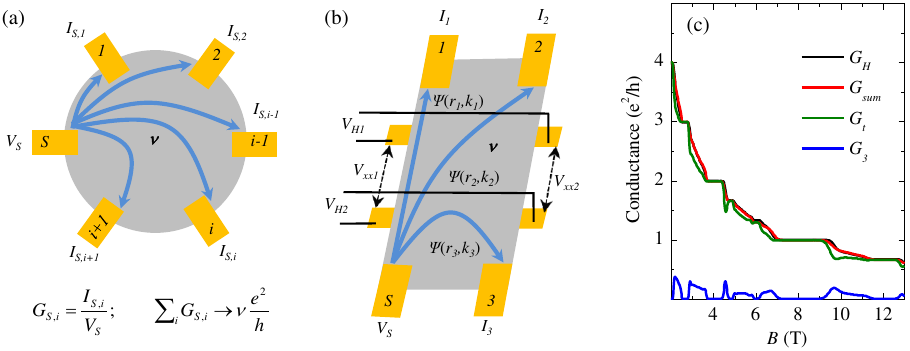}
\caption{\textbf{(a)} Schematic current distribution, with clockwise handedness, among all the contacts connected to compressible QH fluid of filling fraction $\nu$. Voltage $V_S$ is applied at source S and all the other reflection-less contacts with indices $i$ are virtually grounded during output current $I_{S,i}$ measurement. Transmitted conductance at $i^{th}$ contact is measured as $G_{S.i} = I_{S,i}/V_S$. Total sum of conductance of the compressible QH fluid at filling fraction $\nu$ is defined as $\sum_{i}^{} {G_{S,i}}$, which approaches Hall conductance $\nu (e^2/h)$. \textbf{(b)} Schematic measurement setup showing measuring parameters for total conductance study under clockwise handedness/chirality. The wavefunction $\Psi(r_i,k_i)$ represents a chiral trajectory/orbit form source S to $i^{th}$ contact. \textbf{(c)} Plot of conductances $G_t$, $G_3$ and $G_H$ with magnetic field at 30 mK temperature for low density sample. Total sum of conductance $G_{sum} = G_t + G_3$ almost overlaps with the black line of Hall conductance $G_H$.} \label{Fig.1}
  \end{figure*}

In this work, we focus on the transmitted conductance measurements between a source and the reflection-less contacts connected to the compressible QH fluid with filling fraction $\nu$ as shown in Fig.\ref{Fig.1}a. The injected current from source S will be distributed among all contacts and corresponding transmitted conductances ($G_{S,i}$ at $i^{th}$ contact) are measured simultaneously. We show that the total sum of the transmitted conductances approaches the Hall conductance limit $G_H=\nu (e^2/h)$, i.e.$\sum_{i}^{} {G_{S,i}} \xrightarrow{}\nu (e^2/h)$ for all values of $\nu$ of compressible QH fluids. We have established chiral transport in compressible fluids by floating contact measurements. The said sum rule has emerged as a conduction law of chiral metal phase in compressible QH fluids.

For simultaneous measurements of Hall conductance and transmitted conductances, a topologically equivalent Hall bar device is used, as schematically shown in Fig.\ref{Fig.1}b. The Hall bar device is fabricated by UV-photolithography using a one sided modulation doped GaAs/AlGaAs heterostructure grown by molecular beam epitaxy. The transport measurements of the device are carried out using standard Lock-in technique with ac voltage excitation of 25.8 $\mu$V, 17 Hz in a dilution refrigerator with base temperature of 7 mK, where electron temperature is achieved about 30 mK. Output currents are measured by current to voltage pre-amplifiers of gain $10^7$ V/A. Initially, measurements are performed under dark condition at carrier density $\sim$ $2 \times 10^{11}\,\mathrm{cm}^{-2}$ and low-temperature electron mobility $2.70 \times 10^6\,\mathrm{cm^2/Vs}$. The device is also studied at higher density $2.95 \times 10^{11}\,\mathrm{cm}^{-2}$ and mobility $2.98 \times 10^6\,\mathrm{cm^2/Vs}$ after light illumination using a GaAs LED at 3 K \cite{NATHAN1986_persistent}. Detailed characterization of the devices are presented in Supplemental Materials (SM) \cite{supp}. In this study, reflection-less contacts with low contact resistance are essential to measure the transmitted conductance accurately \cite{datta_1995} (see SM \cite{supp}). Furthermore, there must not be any connecting arms to the contacts that can diminish transmitted conductance values through current reflection within the arms \cite{Jain1990_NL_PRL}. 

The single source S is excited with voltage $V_S$ and output current $I_1$, $I_2$ and $I_3$ are measured at the reflection-less contacts 1, 2 and 3 respectively (Fig.\ref{Fig.1}b). Corresponding transmitted conductances ($G_i$) are estimated from the ratio $I_i/V_S$ by series resistance  correction (see SM  \cite{supp}). We define transmitted conductance across the channel of length 200 $\mu$m in the chirality direction as $G_t = G_1 +G_2$. The transmitted conductance $G_3$ across the width of the Hall bar is measured from the current $I_3$. The sum of the transmitted current $I_1+I_2$ across the channel generates Hall voltages $V_{H1}$ and $V_{H2}$; and longitudinal voltages $V_{xx1}$ and $V_{xx2}$. Correspondingly Hall conductance $G_H = \nu (e^2/h)$ is determined independently by four-terminal measurements as $G_H = (I_1+I_2)/V_{H1}$ in nearly uniform Hall bar device \cite{BK2004_PhysicaE} (see SM \cite{supp}).

\begin{figure*}
\includegraphics[width=1\textwidth]{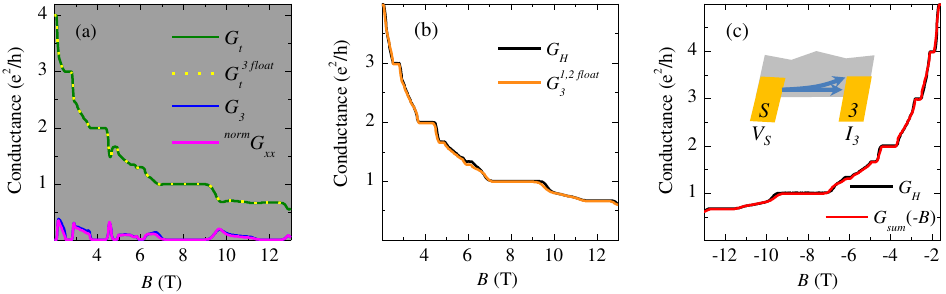}
 \caption{\textbf{(a)} Plot of simultaneously measured conductances $G_t$ and $G_3$ with magnetic field for low density sample at 30 mK. Transmitted conductance $G_{t}^{3 float}$ without measuring $G_3$ is plotted in dotted yellow line. Magenta curve represents the longitudinal conductance ($^{norm}G_{xx}$) normalized to $G_3$. \textbf{(b)} Plot of conductance $G_{3}^{1,2 float}$ under floating condition of contacts 1 and 2, approaches the Hall conductance $G_H$ at 30 mK temperature for low density sample. \textbf{(c)} Plot of total conductance $G_{sum}$(-$B$) for negative magnetic field and comparison with Hall conductance $G_H$ at 30 mK for low density sample. Inset shows shortest current path from source S to contact 3.} \label{Fig.2}
 \end{figure*}
The transmitted conductances $G_t$ and $G_3$ are plotted with magnetic field $B$ along with Hall conductance $G_H$ at temperature 30 mK in Fig.\ref{Fig.1}c. At the QH plateaus, $G_t$ curve coincides with $G_H$ trace, where only  edge modes conduct, resulting in $G_3 = 0$. In the compressible QH region between two QH plateaus, a fraction of distributed bulk current reaches to contact 3 resulting in reduction of $G_t$ value from $G_H$ and the conductance $G_3$ becomes finite as evident in Fig.\ref{Fig.1}c. The total sum of conductances $G_{sum} = G_t+G_3$ almost overlaps with the $G_H$ trace over a wide range of magnetic fields of 2-13 T and filling fraction $\nu = 4-0.6$ in the QH regime. If the compressible fluid is dissipative/percolative, current loops must reduce bulk conductance $G_3$. As a consequence, the total conductance $G_{sum}$, arising from edge and diffusive/percolative bulk transport in compressible fluids, must not approach the Hall conductance $G_H$. Hence, observation of the conductance sum rule indicates suppression of percolation/diffusion in compressible QH fluids.

To verify suppression of percolation or diffusive transport, floating contact conductance measurements are performed. In Fig.\ref{Fig.2}a simultaneously measured conductances $G_t$ and $G_3$ are plotted along with the transmitted conductance $G_{t}^{3 float}$ measured by floating contact 3 for the low density sample. Both the conductances $G_t$ and $G_{t}^{3 float}$ curves nicely overlap with each other in the magnetic field range 2-13 T with root mean square percentage error (RMSPE) $0.5\%$ (see SM \cite{supp}).  
If percolation or diffusive transport is present in the compressible bulk, then a fraction of current reaching contact 3 should reach to contact 1 and 2 when contact 3 is floated (see Fig.\ref{Fig.1}b) and hence $G_{t}^{3 float}$ must be greater than $G_t$ in the compressible regions. But the overlapping of $G_{t}^{3 float}$ and $G_t$ traces in Fig.\ref{Fig.2}a gives null result which confirms the suppression of percolation and diffusive transport in compressible QH bulk fluids. Upon floating contact 3, practically no current reaches to contact 2 and 1 from contact 3 because the current path from contact 3 to 2 and 1 is counter-clockwise, opposite to the chirality direction (Fig.\ref{Fig.1}b).

In other floating contact measurement, when contacts 1 and 2 are floated, all entire current $I_{3}^{1,2 float}$ reaches contact 3 through clockwise chiral paths. Corresponding conductance $G_{3}^{1,2 float}$, estimated from $I_{3}^{1,2 float}/V_s$ (see SM \cite{supp}), recovers the Hall conductance $G_H$ with RMSPE $0.65\%$ as shown in Fig.\ref{Fig.2}b. Therefore, the bulk extended states possess a distinct handedness, where chirality reaches about $100\%$ \cite{meyer2021_chirality} in compressible QH fluids. Hence, the quasi-particles move along clockwise trajectories denoted by extended wave functions  $\Psi{(r_1,k_1)},\Psi{(r_2,k_2)}$ and $\Psi{(r_3,k_3)}$ in the compressible QH regimes as shown in Fig.\ref{Fig.1}b, where quasi-particle exchange is suppressed opposite to the chirality direction.



\begin{figure*}
\includegraphics[width=1\textwidth]{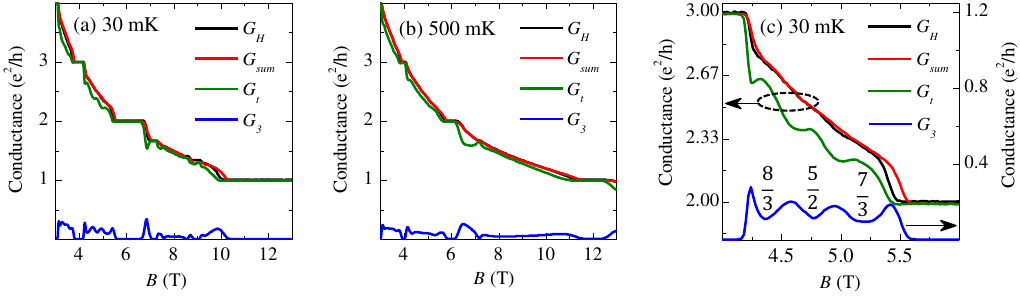}
 \caption{\textbf{(a)} Plot of all conductances with magnetic field for high density sample at 30 mK. \textbf{(b)} Plot of conductances with magnetic field for high density sample at 500 mK. \textbf{(c)} Plot of all conductances with magnetic field for high density sample at 30 mK within filling fraction range $2-3$. Weak $8/3$, $5/2$ and $7/3$ QH states are marked at the deeps in the $G_3$ trace.} \label{Fig.3}
 \end{figure*}
As dissipation within the chiral bulk states is suppressed, the origin of finite longitudinal resistance needs to be well understood in compressible QH fluids. At each contact of a Hall bar device, total incoming current is partitioned according to the conductance ratio of the edge modes and the bulk states, which causes electro-chemical potential equilibration at the contacts \cite{Jain1990_NL_PRL}. Longitudinal voltages are measured from the electro-chemical potentials of the contacts, hence longitudinal resistance ($R_{xx}$) in QH system might be arising from geometric consequence not from dissipation in the bulk (see SM \cite{supp}). From $R_{xx}$ value, the longitudinal (bulk) conductance $G_{xx}$ is calculated. Moreover, $G_3$ represents the transport of pure bulk extended states of partially filled LL. In Fig.\ref{Fig.2}a, $G_{xx}$ is normalized ($^{norm}G_{xx}$) to match the $G_3$ curve by a scaling factor. The matching of the traces confirms that both the conductances $G_{xx}$ and $G_3$ correspond to bulk conductivity $\sigma_{xx}$ (see SM \cite{supp}).

Since dissipation within the chiral bulk states is suppressed, the conduction law of compressible QH fluids might be related to the sum rule of conductance. To explore the sum rule of conductance, chiral bulk transport is studied by reversing the magnetic field, where the chiral current from source S travels a short distance of 20 $\mu$m to reach contact 3 as shown in the inset of Fig.\ref{Fig.2}c. Interestingly, total sum of conductance $G_{sum}(-B)$ approaches the Hall conductance $G_H(-B) = G_H(+B)$ trace for different sample geometry under negative magnetic fields, as shown in Fig.\ref{Fig.2}c (see also SM \cite{supp}). Chiral bulk transport is also studied at a higher carrier density of the 2DEG. Similar to Fig.\ref{Fig.1}c, the conductances are plotted in Fig.\ref{Fig.3}a at 30 mK temperature, where the total sum of conductance $G_{sum}$ approaches the Hall conductance $G_H$. Additionally, the experiment is performed at higher temperatures (see SM \cite{supp}) and the plots at 500 mK in Fig.\ref{Fig.3}b again confirm the approaching of $G_{sum}$ to the $G_H$ limit over a wide range of magnetic field and filling fractions. 

\begin{figure*}
\includegraphics[width=1\textwidth]{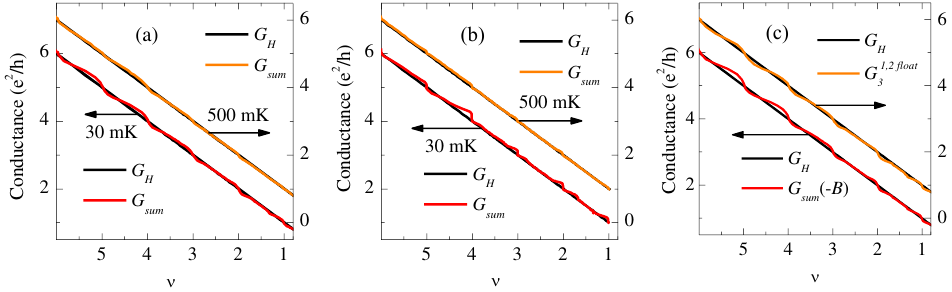}
 \caption{\textbf{(a)} Plot of total conductance $G_{sum}$ and Hall conductance $G_H$ with filling fraction $\nu$ at temperatures 30 mK and 500 mK for low density sample. \textbf{(b)} Plot of $G_{sum}$ and $G_H$ with filling fraction $\nu$ at temperatures 30 mK and 500 mK for high density sample. \textbf{(c)} Plot of total conductances $G_{sum}$(-$B$) and $G_{3}^{1,2float}$ with filling fraction $\nu$ at 30 mK for low density sample.}\label{Fig.4}
\end{figure*}


In the filling fraction range 2 to 3, indication of weak $8/3$, $5/2$ and $7/3$ QH states as the deeps in the conductance $G_3$ trace are obtained in Fig.\ref{Fig.3}c at 30 mK temperature (see SM\cite{supp} for longitudinal resistance $R_{xx}$). In this regime ($B = 4-6 $ T), Coulomb energy modulates the Fermi surface of the system, and, interestingly, the total sum of conductance $G_{sum}$ approaches the $G_H$ trace as well. Therefore, the sum rule of conductance is interaction independent. The result is consistent with the recent observation of the universal Hall response in the interacting atomic system \cite{giamarchi2023_science, Giamarchi2019_prl}.


Goodness of approaching the total sum of conductance $G_{sum}$ to Hall conductance $G_H$ is checked by plotting the conductances in Fig.\ref{Fig.4}(a-c) with the filling fraction $\nu$ derived from Hall conductance $G_H$. To check the goodness, we estimate RMSPE from the data points of the two traces (see SM \cite{supp}).
In Fig.\ref{Fig.4}a, $G_{sum}$ (red line) for low density sample approaches $G_H$ (black line) at 30 mK temperature with RMSPE = $1.44~\%$. At 500 mK temperature, $G_{sum}$ (orange line) for low density sample approaches black line of $G_H$ with RMSPE = $0.65~\%$. For the high density sample, the conductances are plotted in  Fig.\ref{Fig.4}b for temperature 30 mK and 500 mK. At 30 mK and 500 mK temperatures, total sum of conductance $G_{sum}$ approaches $G_H$ with RMSPE = $2.09~\%$ and $0.73~\%$ respectively. In  Fig.\ref{Fig.4}c the conductances $G_{3}^{1,2float}$ and $G_{sum}(-B)$ are approaching Hall conductance $G_H$ with RMSPE = $0.65~\%$ and $~1.68\%$ respectively. The low values of RMSPE in all the plots justify the methodology of series resistance correction (see SM\cite{supp}). In these plots, each incompressible QH plateau is represented by a single point, and rest of the region represents compressible QH regime between two QH plateaus. Hence, all the plots capture the transport property of the compressible QH fluids. The results prove that the sum rule of conductance is universal since it is independent of sample geometry, contact configuration, material quality, magnetic field, temperatures, and quasi-particle interaction. Hence, the sum rule is the fundamental conductance law of compressible QH fluids.

To understand the origin of suppression of dissipative transport with distinct handedness in compressible QH fluids, we consider screening of disorder in the bulk extended states and impact of time reversal symmetry breaking under applied high magnetic fields. The high density of states (DoS) in the partially-filled LL enables screening of external potential \cite{Gerhardts_nps_PhysRevB.38.4218}, such screening produces edge reconstruction \cite{Chklovskii_PhysRevB.46.4026}. The same extended states should also be capable of screening for internal-disorder potential. In high mobility $(> 1\times10^6$ cm$^2/$Vs) 2DEG, average distance $d$ between the local impurities is larger than the magnetic length $l_B = (\sqrt{\hbar/eB})$. In such a condition $d > l_B$, if screening length $l_s$ of the disorder potentials is greater than the magnetic length $l_B$, carrier transport occurs through percolation \cite{trugman1983localization, Percolation_J_T_Chalker_1988}. The screening length $l_s$ can be determined \cite{Ando_RevModPhys.54.437, Dis_screen_PhysRevB.41.1042} as 
\begin{equation}
l_s \approx 
\sqrt{\frac{a\,\epsilon}{4\pi e^2 \left(\frac{dn}{d\mu}\right)}}
\end{equation}
where $a$ is the quantum well width, $\epsilon$ is the dielectric constant of GaAs and $dn/d\mu$ is the DoS at the Fermi energy. If we consider broadening of LL ($\Gamma$), then average DoS of partially filled LL becomes ($eB/h\Gamma$) and correspondingly the ratio $l_s/l_B$ becomes
\begin{equation}
\frac{l_s}{l_B} \approx 
\sqrt{\frac{a\,\epsilon\,\Gamma}{2e^{2}}}
\end{equation}
Putting the values of the parameters $a = 50$ nm, $\epsilon = 12.8 \epsilon_0$ and $\Gamma = 1 $meV \cite{Moty_PRB_DOS}, we find the ratio $l_s/l_B \approx 0.13$. Since $l_s < l_B$ in the 2DEG used, there are no saddle points supporting the percolation process in the compressible QH fluids.

Under the condition $l_s < l_B$, the potential profile of the screened disorder effectively becomes the potential of a point disorder. Although such point disorders facilitate momentum relaxation and resistance at zero magnetic field \cite{DasSarma2015_sR}, the time reversal symmetry breaking at high magnetic fields confirms that no back-scattered states are available at the same locations of the forward moving states. As a result, quasi-particles move in the extended states unidirectionally avoiding local impurities, which gives a new paradigm of conduction within the bulk extended states with suppressed dissipation. Therefore, time reversal symmetry breaking in a flat potential landscape with point disorder in compressible QH fluids results in “chiral metal phase”. We describe this as a distinct phase because it is defined by handedness transport in a gapless system, maintained by the hierarchy of length scales ($l_s < l_B$) and broken time-reversal symmetry. This chiral metal phase is distinctly different from chiral metal formed by parallel edge states in coupled multi-layered 2DEGs \cite{Fisher_PRL1996_chiral_state, Fisher_PRB1997_Chiral_metal, Druist_PRL1998_Chiral_metal,Chalker_PRB1999_Chiral_metal}.

To examine the phase boundary, we consider modulation of screening length in the QH regime. Within incompressible QH plateaus, low DoS at the Fermi energy cannot screen the disorder potential, resulting in $l_s >> l_B$. At the interface of incompressible and compressible regions, the moderate DoS at the Fermi energy might not perfectly screen the disorder potential ($l_s > l_B$), which could favor percolation \cite{trugman1983localization, Percolation_J_T_Chalker_1988}. Within the compressible regions, high DoS at the Fermi energy efficiently screen the disorder potential ($l_s/l_B \sim 0.13 < 1$) that gives chiral metal phase as discussed earlier. Therefore, at the interface of incompressible and compressible regions, percolation might induce small deviation of $G_{sum }$ from $G_H$. Indeed, systematic deviations are seen around the points representing QH plateaus in Fig.\ref{Fig.4}a-b at 30 mK, but such deviations are suppressed possibility due to thermal activation and averaging at 500 mK resulting in lower RMSPE. Lowering of RMSPE with increasing temperature is consistent with the concept of chiral metal phase as it survives at elevated temperatures.

Our findings confirm that the QH fluid is always chiral irrespective of the compressible and incompressible regions at high magnetic fields. However, at zero magnetic field Drude dissipative transport is well established to be isotropic. Hence, our work opens a question about the isotropic to chiral metal phase transition with increasing magnetic field. This chiral metal phase might pave a way for future chiral electronics in semiconductor systems.

When a 2DEG lies in a gapless state, any observed longitudinal resistance is traditionally attributed to Ohmic bulk dissipation. Our work, however, demonstrates a fundamental paradigm shift: a gapless compressible fluid manifests chiral transport with suppressed dissipation under the interplay of short-range screening ($l_s < l_B$) and broken time-reversal symmetry (see SM\cite{supp} for a summary of significance). In such a chiral metal, the longitudinal resistance is entirely non-local, originating from the equilibration of electrochemical potentials at the contacts rather than dissipation within the bulk. Our framework provides a clean blueprint for analyzing transport in other gapless chiral systems, such as Weyl/Dirac semimetal thin films, anomalous Hall systems, thin-film topological insulators, chiral superconductors etc.

\section{Acknowledgements}
Authors thank Deepak Dhar, Krishanu Roychowdhury, Sachindra Nath Karmakar, Pratap Roychowdhury, Chandan Mazumdar, Satyaki Bhattacharya, Arti Garg, 
Steven H. Simon, and Gwendal Feve for insightful discussions and valuable suggestions. BK thanks DAE for funding through plan project.

\section{Data availability}
All the data supporting the findings of this study are available upon reasonable request.


\bibliography{bibfile.bib}

@misc{supp,
  title = {},
  author = {},
  year = {},
  note = {See the Supplemental Material (SM) at XXXXXXXXXXX for detailed discussions.
 },
  url = {}
}

@article{Klitzing1980_PRL,
  title = {New {Method} for {High-Accuracy} {Determination} of the {Fine-Structure Constant Based} on {Quantized Hall Resistance}},
  author = {Klitzing, K. v. and Dorda, G. and Pepper, M.},
  journal = {Phys. Rev. Lett.},
  volume = {45},
  issue = {6},
  pages = {494--497},
  numpages = {0},
  year = {1980},
  month = {Aug},
  publisher = {American Physical Society},
  doi = {10.1103/PhysRevLett.45.494},
  url = {https://link.aps.org/doi/10.1103/PhysRevLett.45.494}
}

@article{Tsui1982_PRL,
  title = {{Two-Dimensional Magnetotransport} in the {Extreme Quantum Limit}},
  author = {Tsui, D. C. and Stormer, H. L. and Gossard, A. C.},
  journal = {Phys. Rev. Lett.},
  volume = {48},
  issue = {22},
  pages = {1559--1562},
  numpages = {0},
  year = {1982},
  month = {May},
  publisher = {American Physical Society},
  doi = {10.1103/PhysRevLett.48.1559},
  url = {https://link.aps.org/doi/10.1103/PhysRevLett.48.1559}
}

@book{Prange1987, 
place={New York, NY}, 
 title={The Quantum Hall Effect}, 
 ISBN={9781468404999}, 
 ISSN={0938-037X}, 
 url={http://dx.doi.org/10.1007/978-1-4684-0499-9},
 DOI={10.1007/978-1-4684-0499-9}, 
 journal={Graduate Texts in Contemporary Physics}, 
 editors={Richard E. Prange and Steven M. Girvin},
 publisher={editors Richard E. Prange and Steven M. Girvin, Springer US}, 
 year={1987} 
 }

@article{Laughlin1983_PRL,
  title = {{Anomalous Quantum Hall Effect: An Incompressible Quantum Fluid} with {Fractionally Charged Excitations}},
  author = {Laughlin, R. B.},
  journal = {Phys. Rev. Lett.},
  volume = {50},
  issue = {18},
  pages = {1395--1398},
  numpages = {0},
  year = {1983},
  month = {May},
  publisher = {American Physical Society},
  doi = {10.1103/PhysRevLett.50.1395},
  url = {https://link.aps.org/doi/10.1103/PhysRevLett.50.1395}
}

@article{Jain_PhysRevLett.63.199,
  title = {Composite-fermion approach for the fractional quantum {Hall} effect},
  author = {Jain, J. K.},
  journal = {Phys. Rev. Lett.},
  volume = {63},
  issue = {2},
  pages = {199--202},
  numpages = {0},
  year = {1989},
  month = {Jul},
  publisher = {American Physical Society},
  doi = {10.1103/PhysRevLett.63.199},
  url = {https://link.aps.org/doi/10.1103/PhysRevLett.63.199}
}

@book{Jain2007_book, 
place={Cambridge}, 
title={Composite Fermions}, 
publisher={Cambridge University Press}, 
author={Jain, Jainendra K.}, 
year={2007}
}

@article{Halperin1982_PRB,
  title = {Quantized {Hall} conductance, current-carrying edge states, and the existence of extended states in a two-dimensional disordered potential},
  author = {Halperin, B. I.},
  journal = {Phys. Rev. B},
  volume = {25},
  issue = {4},
  pages = {2185--2190},
  numpages = {0},
  year = {1982},
  month = {Feb},
  publisher = {American Physical Society},
  doi = {10.1103/PhysRevB.25.2185},
  url = {https://link.aps.org/doi/10.1103/PhysRevB.25.2185}
}

@article{Wen1990_PRB,
  title = {Chiral {Luttinger} liquid and the edge excitations in the fractional quantum {Hall} states},
  author = {Wen, X. G.},
  journal = {Phys. Rev. B},
  volume = {41},
  issue = {18},
  pages = {12838--12844},
  numpages = {0},
  year = {1990},
  month = {Jun},
  publisher = {American Physical Society},
  doi = {10.1103/PhysRevB.41.12838},
  url = {https://link.aps.org/doi/10.1103/PhysRevB.41.12838}
}

@article{Buttiker1986-PRL,
  title = {Four-Terminal Phase-Coherent Conductance},
  author = {B\"uttiker, M.},
  journal = {Phys. Rev. Lett.},
  volume = {57},
  issue = {14},
  pages = {1761--1764},
  numpages = {0},
  year = {1986},
  month = {Oct},
  publisher = {American Physical Society},
  doi = {10.1103/PhysRevLett.57.1761},
  url = {https://link.aps.org/doi/10.1103/PhysRevLett.57.1761}
}

@article{buttiker1988_backscattering,
  title = {Absence of backscattering in the quantum {Hall} effect in multiprobe conductors},
  author = {B\"uttiker, M.},
  journal = {Phys. Rev. B},
  volume = {38},
  issue = {14},
  pages = {9375--9389},
  numpages = {0},
  year = {1988},
  month = {Nov},
  publisher = {American Physical Society},
  doi = {10.1103/PhysRevB.38.9375},
  url = {https://link.aps.org/doi/10.1103/PhysRevB.38.9375}
}

@article{Maiti2021_PRB,
  title = {Temperature-dependent equilibration of spin orthogonal quantum {Hall} edge modes},
  author = {Maiti, Tanmay and Agarwal, Pooja and Purkait, Suvankar and Sreejith, G. J. and Das, Sourin and Biasiol, Giorgio and Sorba, Lucia and Karmakar, Biswajit},
  journal = {Phys. Rev. B},
  volume = {104},
  issue = {8},
  pages = {085304},
  numpages = {7},
  year = {2021},
  month = {Aug},
  publisher = {American Physical Society},
  doi = {10.1103/PhysRevB.104.085304},
  url = {https://link.aps.org/doi/10.1103/PhysRevB.104.085304}
}

@article{Sirt2025_APL,
    author = {Sirt, S. and Umansky, V. Y. and Siddiki, A. and Ludwig, S.},
    title = {Transition from edge- to bulk-currents in the quantum {Hall} regime},
    journal = {Applied Physics Letters},
    volume = {126},
    number = {24},
    pages = {243101},
    year = {2025},
    month = {06},
    issn = {0003-6951},
    doi = {10.1063/5.0275599},
    url = {https://doi.org/10.1063/5.0275599},
}

@article{Sirt2025_NJP,
doi = {10.1088/1367-2630/adfc05},
url = {https://doi.org/10.1088/1367-2630/adfc05},
year = {2025},
month = {aug},
publisher = {IOP Publishing},
volume = {27},
number = {8},
pages = {083507},
author = {Sirt, Serkan and Kamm, Matthias and Umansky, Vladimir Y and Ludwig, Stefan},
title = {Nature of current flow in the regime of the quantum {Hall} effect},
journal = {New Journal of Physics},
}

@article{Maity2020_PRL,
  title = {{Magnetic-Field-Dependent Equilibration} of {Fractional Quantum Hall Edge Modes}},
  author = {Maiti, Tanmay and Agarwal, Pooja and Purkait, Suvankar and Sreejith, G. J. and Das, Sourin and Biasiol, Giorgio and Sorba, Lucia and Karmakar, Biswajit},
  journal = {Phys. Rev. Lett.},
  volume = {125},
  issue = {7},
  pages = {076802},
  numpages = {6},
  year = {2020},
  month = {Aug},
  publisher = {American Physical Society},
  doi = {10.1103/PhysRevLett.125.076802},
  url = {https://link.aps.org/doi/10.1103/PhysRevLett.125.076802}
}

@article{sabo2017_nPhy,
  title={Edge reconstruction in fractional quantum {Hall} states},
  author={Sabo, R. and Gurman, I. and Rosenblatt, A. and Lafont, F. and Banitt, D. and Park, J. and Heiblum, M. and Gefen, Y. and Umansky, V. and Mahalu, D.},
  journal={Nature Physics},
  volume={13},
  number={5},
  pages={491},
  year={2017},
  publisher={Nature Publishing Group},
  doi={10.1038/nphys4010},
  url={https://doi.org/10.1038/nphys4010}
}

@article{Nakamura2023_prl,
  title = {{Half-Integer Conductance Plateau} at the {$\ensuremath\nu=2/3$ Fractional Quantum Hall State} in a {Quantum Point Contact}},
  author = {Nakamura, J. and Liang, S. and Gardner, G. C. and Manfra, M. J.},
  journal = {Phys. Rev. Lett.},
  volume = {130},
  issue = {7},
  pages = {076205},
  numpages = {7},
  year = {2023},
  month = {Feb},
  publisher = {American Physical Society},
  doi = {10.1103/PhysRevLett.130.076205},
  url = {https://link.aps.org/doi/10.1103/PhysRevLett.130.076205}
}

@article{Son1991_PRB,
  title = {Nonequilibrium distribution of edge and bulk current in a quantum {Hall} conductor},
  author = {van Son, P. C. and de Vries, F. W. and Klapwijk, T. M.},
  journal = {Phys. Rev. B},
  volume = {43},
  issue = {8},
  pages = {6764--6767},
  numpages = {0},
  year = {1991},
  month = {Mar},
  publisher = {American Physical Society},
  doi = {10.1103/PhysRevB.43.6764},
  url = {https://link.aps.org/doi/10.1103/PhysRevB.43.6764}
}

@article{Hirai1994_PRB,
  title = {Ratio between edge and bulk currents in the quantum {Hall} effect},
  author = {Hirai, H. and Komiyama, S.},
  journal = {Phys. Rev. B},
  volume = {49},
  issue = {19},
  pages = {14012--14015},
  numpages = {0},
  year = {1994},
  month = {May},
  publisher = {American Physical Society},
  doi = {10.1103/PhysRevB.49.14012},
  url = {https://link.aps.org/doi/10.1103/PhysRevB.49.14012}
}

@article{Klitzing2001_dist,
title = {Hall potential profiles in the quantum {Hall} regime measured by a scanning force microscope},
journal = {Physica B: Condensed Matter},
volume = {298},
number = {1},
pages = {562-566},
year = {2001},
note = {International Conference on High Magnetic Fields in Semiconductors},
issn = {0921-4526},
doi = {https://doi.org/10.1016/S0921-4526(01)00383-0},
url = {https://www.sciencedirect.com/science/article/pii/S0921452601003830},
author = {E. Ahlswede and P. Weitz and J. Weis and K. {von Klitzing} and K. Eberl},
}

@article{li2005scaling_PhysRevLett.94.206807,
  title = {Scaling and Universality of Integer Quantum Hall Plateau-to-Plateau Transitions},
  author = {Li, Wanli and Cs\'athy, G. A. and Tsui, D. C. and Pfeiffer, L. N. and West, K. W.},
  journal = {Phys. Rev. Lett.},
  volume = {94},
  issue = {20},
  pages = {206807},
  numpages = {4},
  year = {2005},
  month = {May},
  publisher = {American Physical Society},
  doi = {10.1103/PhysRevLett.94.206807},
  url = {https://link.aps.org/doi/10.1103/PhysRevLett.94.206807}
}

@article{Shahar1997_PRL,
  title = {{A Different View} of the {Quantum Hall} {Plateau}-to-{Plateau Transitions}},
  author = {Shahar, D. and Tsui, D. C. and Shayegan, M. and Shimshoni, E. and Sondhi, S. L.},
  journal = {Phys. Rev. Lett.},
  volume = {79},
  issue = {3},
  pages = {479--482},
  numpages = {0},
  year = {1997},
  month = {Jul},
  publisher = {American Physical Society},
  doi = {10.1103/PhysRevLett.79.479},
  url = {https://link.aps.org/doi/10.1103/PhysRevLett.79.479}
}

@article{Jain1990_NL_PRL,
  title = {New resistivity for high-mobility quantum {Hall} conductors},
  author = {McEuen, P. L. and Szafer, A. and Richter, C. A. and Alphenaar, B. W. and Jain, J. K. and Stone, A. D. and Wheeler, R. G. and Sacks, R. N.},
  journal = {Phys. Rev. Lett.},
  volume = {64},
  issue = {17},
  pages = {2062--2065},
  numpages = {0},
  year = {1990},
  month = {Apr},
  publisher = {American Physical Society},
  doi = {10.1103/PhysRevLett.64.2062},
  url = {https://link.aps.org/doi/10.1103/PhysRevLett.64.2062}
}

@article{Wang1992_PRB,
  title = {Measurements and modeling of nonlocal resistance in the fractional quantum {Hall} effect},
  author = {Wang, J. K. and Goldman, V. J.},
  journal = {Phys. Rev. B},
  volume = {45},
  issue = {23},
  pages = {13479--13487},
  numpages = {0},
  year = {1992},
  month = {Jun},
  publisher = {American Physical Society},
  doi = {10.1103/PhysRevB.45.13479},
  url = {https://link.aps.org/doi/10.1103/PhysRevB.45.13479}
}

@article{Purkait2024_PRB1,
  title = {Edge reconstruction of a compressible quantum Hall fluid in the filling fraction range 1/3 to 2/3},
  author = {Purkait, Suvankar and Maiti, Tanmay and Agarwal, Pooja and Sahoo, Suparna and G J, Sreejith and Das, Sourin and Biasiol, Giorgio and Sorba, Lucia and Karmakar, Biswajit},
  journal = {Phys. Rev. B},
  volume = {110},
  issue = {24},
  pages = {245309},
  numpages = {8},
  year = {2024},
  month = {Dec},
  publisher = {American Physical Society},
  doi = {10.1103/PhysRevB.110.245309},
  url = {https://link.aps.org/doi/10.1103/PhysRevB.110.245309}
}

@article{Purkait2025_PRB2,
  title = {Collapse of edge reconstruction in compressible quantum {Hall} fluid within filling fraction range $\frac{2}{3}$ to 1},
  author = {Purkait, Suvankar and Maiti, Tanmay and Agarwal, Pooja and Sahoo, Suparna and Biasiol, Giorgio and Sorba, Lucia and Karmakar, Biswajit},
  journal = {Phys. Rev. B},
  volume = {112},
  issue = {11},
  pages = {115306},
  numpages = {10},
  year = {2025},
  month = {Sep},
  publisher = {American Physical Society},
  doi = {10.1103/txxb-47vd},
  url = {https://link.aps.org/doi/10.1103/txxb-47vd}
}

@article{wei1994_PRB_scaling,
  title = {Current scaling in the integer quantum {Hall} effect},
  author = {Wei, H. P. and Engel, L. W. and Tsui, D. C.},
  journal = {Phys. Rev. B},
  volume = {50},
  issue = {19},
  pages = {14609--14612},
  numpages = {0},
  year = {1994},
  month = {Nov},
  publisher = {American Physical Society},
  doi = {10.1103/PhysRevB.50.14609},
  url = {https://link.aps.org/doi/10.1103/PhysRevB.50.14609}
}

@book{datta_1995, place={Cambridge}, series={Cambridge Studies in Semiconductor Physics and Microelectronic Engineering}, title={Electronic Transport in Mesoscopic Systems}, DOI={10.1017/CBO9780511805776}, publisher={Cambridge University Press}, author={Datta, Supriyo}, year={1995}, collection={Cambridge Studies in Semiconductor Physics and Microelectronic Engineering}
}

@article{giamarchi2023_science,
author = {T.-W. Zhou  and G. Cappellini  and D. Tusi  and L. Franchi  and J. Parravicini  and C. Repellin  and S. Greschner  and M. Inguscio  and T. Giamarchi  and M. Filippone  and J. Catani  and L. Fallani },
title = {Observation of universal {Hall} response in strongly interacting {Fermions}},
journal = {Science},
volume = {381},
number = {6656},
pages = {427-430},
year = {2023},
doi = {10.1126/science.add1969},
URL = {https://www.science.org/doi/abs/10.1126/science.add1969},
}

@article{Giamarchi2019_prl,
  title = {Universal {Hall Response} in {Interacting Quantum Systems}},
  author = {Greschner, Sebastian and Filippone, Michele and Giamarchi, Thierry},
  journal = {Phys. Rev. Lett.},
  volume = {122},
  issue = {8},
  pages = {083402},
  numpages = {6},
  year = {2019},
  month = {Feb},
  publisher = {American Physical Society},
  doi = {10.1103/PhysRevLett.122.083402},
  url = {https://link.aps.org/doi/10.1103/PhysRevLett.122.083402}
}

@article{BK2004_PhysicaE,
title = {The effects of macroscopic inhomogeneities on the magnetotransport properties of the electron gas in two dimensions},
journal = {Physica E: Low-dimensional Systems and Nanostructures},
volume = {24},
number = {3},
pages = {187-210},
year = {2004},
issn = {1386-9477},
doi = {https://doi.org/10.1016/j.physe.2004.03.019},
url = {https://www.sciencedirect.com/science/article/pii/S1386947704001110},
author = {B. Karmakar and M.R. Gokhale and A.P. Shah and B.M. Arora and D.T.N. {de Lang} and A. {de Visser} and L.A. Ponomarenko and A.M.M. Pruisken},
}

@article{oswald1998_PhysicaE,
title = {A new model for the transport regime of the integer quantum {Hall} effect: The role of bulk transport in the edge channel picture},
journal = {Physica E: Low-dimensional Systems and Nanostructures},
volume = {3},
number = {1},
pages = {30-37},
year = {1998},
issn = {1386-9477},
doi = {https://doi.org/10.1016/S1386-9477(98)00215-X},
url = {https://www.sciencedirect.com/science/article/pii/S138694779800215X},
author = {Josef Oswald},
}

@article{meyer2021_chirality,
  title = {Conductance quantization in topological Josephson trijunctions},
  author = {Meyer, Julia S. and Houzet, Manuel},
  journal = {Phys. Rev. B},
  volume = {103},
  issue = {17},
  pages = {174504},
  numpages = {8},
  year = {2021},
  month = {May},
  publisher = {American Physical Society},
  doi = {10.1103/PhysRevB.103.174504},
  url = {https://link.aps.org/doi/10.1103/PhysRevB.103.174504}
}

@article{NATHAN1986_persistent,
title = {Persistent photoconductivity in {AlGaAs}/{GaAs} modulation doped layers and field effect transistors: {A} review},
journal = {Solid-State Electronics},
volume = {29},
number = {2},
pages = {167-172},
year = {1986},
issn = {0038-1101},
doi = {https://doi.org/10.1016/0038-1101(86)90035-3},
url = {https://www.sciencedirect.com/science/article/pii/0038110186900353},
author = {Marshall I. Nathan},
}

@article{Percolation_J_T_Chalker_1988,
doi = {10.1088/0022-3719/21/14/008},
url = {https://doi.org/10.1088/0022-3719/21/14/008},
year = {1988},
month = {may},
publisher = {},
volume = {21},
number = {14},
pages = {2665},
author = {J T Chalker and P D Coddington},
title = {Percolation, quantum tunnelling and the integer {Hall} effect},
journal = {Journal of Physics C: Solid State Physics},
}

@article{trugman1983localization,
  title = {Localization, percolation, and the quantum {Hall} effect},
  author = {Trugman, S. A.},
  journal = {Phys. Rev. B},
  volume = {27},
  issue = {12},
  pages = {7539--7546},
  numpages = {0},
  year = {1983},
  month = {Jun},
  publisher = {American Physical Society},
  doi = {10.1103/PhysRevB.27.7539},
  url = {https://link.aps.org/doi/10.1103/PhysRevB.27.7539}
}

@article{Gerhardts_nps_PhysRevB.38.4218,
  title = {Screening properties of the two-dimensional electron gas in the quantum {Hall} regime},
  author = {Wulf, Ulrich and Gudmundsson, Vidar and Gerhardts, Rolf R.},
  journal = {Phys. Rev. B},
  volume = {38},
  issue = {6},
  pages = {4218--4230},
  numpages = {0},
  year = {1988},
  month = {Aug},
  publisher = {American Physical Society},
  doi = {10.1103/PhysRevB.38.4218},
  url = {https://link.aps.org/doi/10.1103/PhysRevB.38.4218}
}

@article{Dis_screen_PhysRevB.41.1042,
  title = {Disorder, screening, and quantum {Hall} oscillations},
  author = {Esfarjani, Keivan and Glyde, Henry R. and Sa-yakanit, Virulh},
  journal = {Phys. Rev. B},
  volume = {41},
  issue = {2},
  pages = {1042--1053},
  numpages = {0},
  year = {1990},
  month = {Jan},
  publisher = {American Physical Society},
  doi = {10.1103/PhysRevB.41.1042},
  url = {https://link.aps.org/doi/10.1103/PhysRevB.41.1042}
}

@article{Ando_RevModPhys.54.437,
  title = {Electronic properties of two-dimensional systems},
  author = {Ando, Tsuneya and Fowler, Alan B. and Stern, Frank},
  journal = {Rev. Mod. Phys.},
  volume = {54},
  issue = {2},
  pages = {437--672},
  numpages = {0},
  year = {1982},
  month = {Apr},
  publisher = {American Physical Society},
  doi = {10.1103/RevModPhys.54.437},
  url = {https://link.aps.org/doi/10.1103/RevModPhys.54.437}
}

@Article{DasSarma2015_sR,
author={Das Sarma, S.
and Hwang, E. H.},
title={Screening and transport in 2D semiconductor systems at low temperatures},
journal={Scientific Reports},
year={2015},
month={Nov},
day={17},
volume={5},
number={1},
pages={16655},
issn={2045-2322},
doi={10.1038/srep16655},
url={https://doi.org/10.1038/srep16655}
}

@article{Chklovskii_PhysRevB.46.4026,
  title = {Electrostatics of edge channels},
  author = {Chklovskii, D. B. and Shklovskii, B. I. and Glazman, L. I.},
  journal = {Phys. Rev. B},
  volume = {46},
  issue = {7},
  pages = {4026--4034},
  numpages = {0},
  year = {1992},
  month = {Aug},
  publisher = {American Physical Society},
  doi = {10.1103/PhysRevB.46.4026},
  url = {https://link.aps.org/doi/10.1103/PhysRevB.46.4026}
}

@article{Moty_PRB_DOS,
  title = {Direct measurement of the density of states of a two-dimensional electron gas},
  author = {Smith, T. P. and Goldberg, B. B. and Stiles, P. J. and Heiblum, M.},
  journal = {Phys. Rev. B},
  volume = {32},
  issue = {4},
  pages = {2696--2699},
  numpages = {0},
  year = {1985},
  month = {Aug},
  publisher = {American Physical Society},
  doi = {10.1103/PhysRevB.32.2696},
  url = {https://link.aps.org/doi/10.1103/PhysRevB.32.2696}
}

@article{Fisher_PRL1996_chiral_state,
  title = {Chiral Surface States in the Bulk Quantum Hall Effect},
  author = {Balents, Leon and Fisher, Matthew P. A.},
  journal = {Phys. Rev. Lett.},
  volume = {76},
  issue = {15},
  pages = {2782--2785},
  numpages = {0},
  year = {1996},
  month = {Apr},
  publisher = {American Physical Society},
  doi = {10.1103/PhysRevLett.76.2782},
  url = {https://link.aps.org/doi/10.1103/PhysRevLett.76.2782}
}

@article{Fisher_PRB1997_Chiral_metal,
  title = {Transport of surface states in the bulk quantum Hall effect},
  author = {Cho, Sora and Balents, Leon and Fisher, Matthew P. A.},
  journal = {Phys. Rev. B},
  volume = {56},
  issue = {24},
  pages = {15814--15821},
  numpages = {0},
  year = {1997},
  month = {Dec},
  publisher = {American Physical Society},
  doi = {10.1103/PhysRevB.56.15814},
  url = {https://link.aps.org/doi/10.1103/PhysRevB.56.15814}
}

@article{Druist_PRL1998_Chiral_metal,
  title = {Observation of Chiral Surface States in the Integer Quantum Hall Effect},
  author = {Druist, D. P. and Turley, P. J. and Maranowski, K. D. and Gwinn, E. G. and Gossard, A. C.},
  journal = {Phys. Rev. Lett.},
  volume = {80},
  issue = {2},
  pages = {365--368},
  numpages = {0},
  year = {1998},
  month = {Jan},
  publisher = {American Physical Society},
  doi = {10.1103/PhysRevLett.80.365},
  url = {https://link.aps.org/doi/10.1103/PhysRevLett.80.365}
}

@article{Chalker_PRB1999_Chiral_metal,
  title = {Transverse magnetoresistance of the two-dimensional chiral metal},
  author = {Chalker, J. T. and Sondhi, S. L.},
  journal = {Phys. Rev. B},
  volume = {59},
  issue = {7},
  pages = {4999--5002},
  numpages = {0},
  year = {1999},
  month = {Feb},
  publisher = {American Physical Society},
  doi = {10.1103/PhysRevB.59.4999},
  url = {https://link.aps.org/doi/10.1103/PhysRevB.59.4999}
}

\newpage
\clearpage	
\normalsize\clearpage

\begin{onecolumngrid}
		\begin{center}
			{\fontsize{12}{12}\selectfont
				\textbf{Supplemental Material for\\
                  \vspace{3mm} 
``Emergent Chiral Metal Phase in Compressible Quantum Hall Fluids''\\[5mm]}}
			{\normalsize  Suparna Sahoo,$^{1,2}$ Suvankar Purkait,$^{1,2}$ Pooja Agarwal,$^{3}$ Tanmay Maiti,$^{4}$\\ Sourin Das,$^{5}$ Vladimir Umansky,$^{6}$ and Biswajit Karmakar,$^{1,2}$\\[3mm]}
			{\small $^1$\textit{Saha Institute of Nuclear Physics, 1/AF Bidhannagar, Kolkata 700 064, India}\\[0.5mm]}
			{\small $^2$\textit{Homi Bhabha National Institute, Anushaktinagar, Mumbai 400094, India}\\[0.5mm]}
			{\small $^3$\textit{SPEC, CEA, CNRS, Université Paris-Saclay, CEA Saclay, 91191 Gif sur Yvette Cedex, France}\\[0.5mm]}
			{\small $^4$\textit{Department of Physics and Astronomy, Purdue University, West Lafayette, Indiana 47907, USA}\\[0.5mm]}
            {\small $^5$\textit{Department of Physical Sciences, IISER Kolkata, Mohanpur, West Bengal 741246, India}\\[0.5mm]}
            {\small $^6$\textit{Department of Condensed Matter Physics, Weizmann Institute of Science, Rehovot 7610001, Israel}\\[0.5mm]}
			{}
		\end{center}
		
		\newcounter{defcounter}
		\setcounter{defcounter}{0}
		\setcounter{equation}{0}
		\setcounter{page}{1}
		\pagenumbering{roman}
		
		\renewcommand{\thesection}{S\araic{section}}
		
		\newcounter{specialsection}
		\renewcommand{\thespecialsection}{S.\Roman{specialsection}. }
		
        \newcommand{\specialsection}[1]{
			\refstepcounter{specialsection}
			\addcontentsline{stoc}{section}
			{\protect\numberline{\thespecialsection}#1}
			\bigskip
			\noindent
			{\normalsize\bfseries
				\thespecialsection #1\quad}
			\par\medskip
		}



\begin{center}
\specialsection{\,Device Structure} 
\end{center}		
\label{sec:Device Structure}
Optical image of the Hall bar device used in this experiment is shown in Fig.~\ref{Fig_S1} including the measurement parameters. The mesa is defined by UV-photolithography and Ohmic contacts are made by subsequent thermal evaporation of metals Ni/Au-Ge/Ni/Au and rapid thermal annealing. The device is fabricated using a one sided modulation doped GaAs/AlGaAs hetero-structure, where the two-dimensional electron gas (2DEG) resides 83.5 nm below the top surface. The length of the mesa is 200 $\mu$m (contact S to contact 1) and width is 60 $\mu$m. The distance from contact S to contact 3 is 20 $\mu$m approx. The contact arms are 20 $\mu$m wide approximately. Source and detector contacts are placed directly on the mesa having overlap of 20 $\mu$m x 20 $\mu$m approx.

\begin{figure*}[h]
\includegraphics[width=0.45\textwidth]{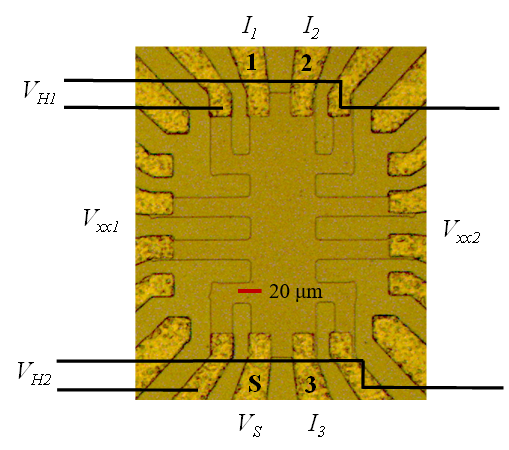}
 \caption{Optical image of the Hall bar device with measurement configuration.}\label{Fig_S1}
\end{figure*}

\begin{center}
	\specialsection{\, Device Characterization under dark condition} 
\end{center}
\label{sec:Characterization_Dark}
At first the transport measurements of the device are performed under dark condition without any light illumination. As shown in Fig.~\ref{Fig_S1}, two Hall voltages $V_{H1}$ and $V_{H2}$, and two longitudinal voltages $V_{xx1}$ and $V_{xx2}$ are measured simultaneously by varying magnetic field 0-13 T at 30 mK temperature. The Hall resistance $R_H$ and longitudinal resistance $R_{xx}$ are calculated as $R_H = V_H/(I_1+I_2)$ and $R_{xx} = V_{xx}/(I_1+I_2)$ respectively.  It is hard to know the exact current distribution in
the sample, but the net transmitted current $I_1+I_2$ determines the Hall and longitudinal resistance. Corresponding two sets of Hall and longitudinal resistances are plotted in Fig.~\ref{Fig_S2}. Nearly overlapping traces of the Hall resistances $R_{H1}$ and $R_{H2}$ confirm the density uniformity ~\cite{BK2004_PhysicaE} of the Hall bar device. The results also prove that the current distribution has the least impact on Hall voltage measurements. Actually, Hall voltage is determined by integrating the local Hall voltage across the width of the sample. Therefore, Hall voltage is independent of the current distribution as long as the sample possesses density uniformity. At 30 mK temperature the carrier density $n$ and mobility $\mu$ of the 2DEG are determined as $n\sim$ $2 \times 10^{11}\,\mathrm{cm}^{-2}$ and $\mu \sim 2.70 \times 10^6\,\mathrm{cm^2/Vs}$ respectively under dark condition. Two longitudinal resistances $R_{xx1}$ and $R_{xx2}$ also show similar behavior with magnetic field variation.

\begin{figure*}[h]
\includegraphics[width=0.45\textwidth]{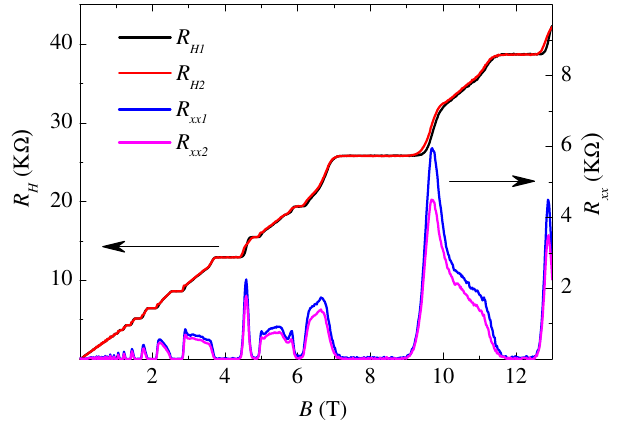}
 \caption{Plot of quantum Hall data at 30 mK under dark condition.}\label{Fig_S2}
\end{figure*}

\begin{center}
	\specialsection{\, Device Characterization after light illumination} 
\end{center}
\label{sec:Characterization_illumination}
After light illumination from a GaAs LED at 3~K temperature, the carrier density and mobility of the 2DEG are expected to be higher, and persistent photoconductivity holds the carriers at low temperatures~\cite{NATHAN1986_persistent}. All measurements are repeated after light illumination, and the corresponding Hall and longitudinal resistance traces are plotted in Fig.~\ref{Fig_S3}. Overlapping of two Hall traces and similarly overlapping of two longitudinal resistance traces confirm the density homogeneity~\cite{BK2004_PhysicaE} of the device. The carrier density and mobility are found to be $n \sim 2.95\times10^{11}~\mathrm{cm}^{-2}$ and $\mu \sim 2.98\times10^{6}~\mathrm{cm}^{2}/\mathrm{Vs}$, respectively, at the lowest temperature. In this high-mobility condition, signatures of FQH states like $8/3$, $5/2$, and $7/3$ are obtained as the dips in the $R_{xx}$ trace between 4--6~T.

\begin{figure*}[h]
\includegraphics[width=0.45\textwidth]{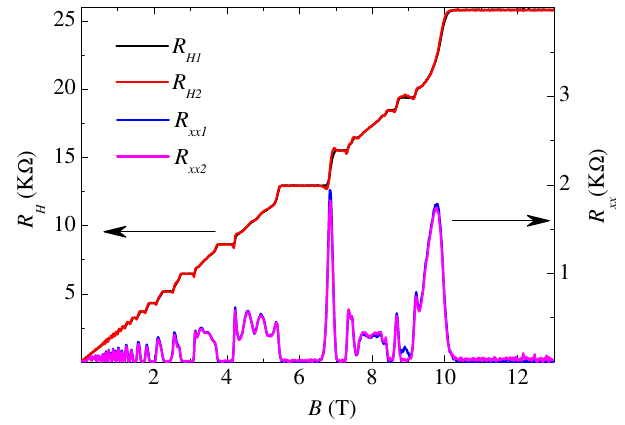}
 \caption{Plot of quantum Hall data at 30 mK after light illumination.}\label{Fig_S3}
\end{figure*}

\begin{center}
	\specialsection{\, Comparison of quantum Hall data at temperatures 30 mK and 500 mK under dark condition} 
\end{center}
\label{sec:Hall data 30 & 500 mK_Dark}
Comparison of Hall resistance $R_{H1}$ and longitudinal resistance $R_{xx1}$ at 30 mK and 500 mK under dark condition is shown in Fig.~\ref{Fig_S4}. At higher temperature, quantum Hall plateau widths become narrower. Accordingly the $R_{xx1}$ curve becomes smoother at 500 mK temperature. The different traces of Hall resistance suggest significant change of Hall conductance $G_H=1/R_{H1}$ with magnetic field at two different temperatures. These estimated Hall conductance $G_H$ is used to study universal total sum of conductance.
\begin{figure*}[h]
\includegraphics[width=0.45\textwidth]{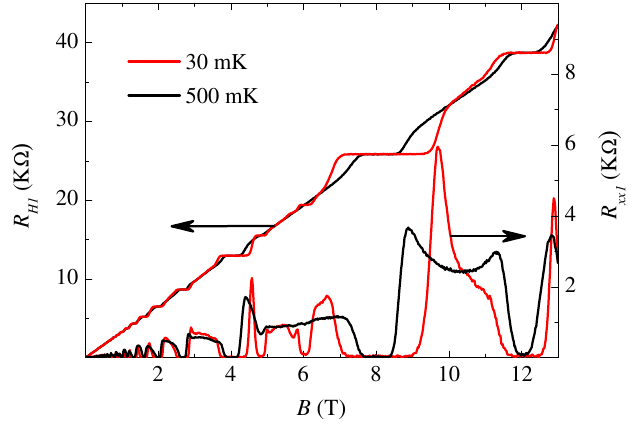}
 \caption{Plot of quantum Hall data at 30 mK and 500 mK under dark condition.}\label{Fig_S4}
\end{figure*}
\\
\\
\begin{center}
	\specialsection{\, Comparison of quantum Hall data at temperatures 30 mK and 500 mK after light illumination} 
\end{center}
\label{sec:Hall data 30 & 500 mK_Illumination}
Fig.~\ref{Fig_S5} represents the comparison of Hall resistance $R_{H1}$ and longitudinal resistance $R_{xx1}$ for two different temperatures 30 mK and 500 mK after carrier injection by light illumination. At higher temperature the prominent FQH plateaus are disappeared and integer quantum Hall plateaus become narrower. Similarly the structures in $R_{xx1}$ become much weaker at higher temperature. From these Hall traces Hall conductance $G_H = 1/R_{H1}$ is estimated and used for universal total conductance study.
\begin{figure*}[h]
\includegraphics[width=0.45\textwidth]{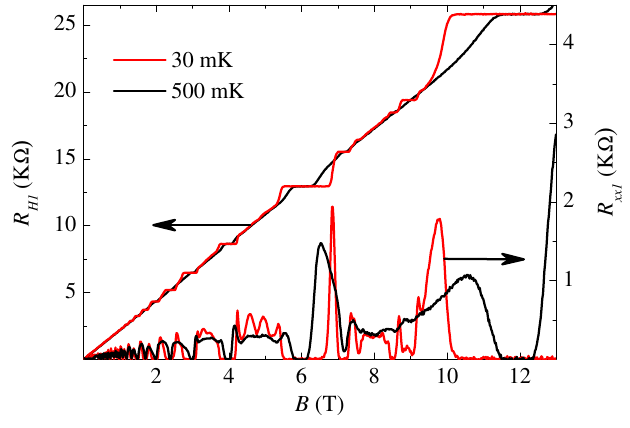}
 \caption{Plot of quantum Hall data at 30 mK and 500 mK after light illumination.}\label{Fig_S5}
\end{figure*}

\begin{center}
	\specialsection{\, Systematic series resistance correction to the measured conductance data} 
\end{center}
\label{sec:series resistance correction}
Transmitted conductances are measured by two-terminal configuration as shown in Fig.~\ref{Fig_S1} and Fig.~\ref{Fig.1}b of main text. For transmitted conductance measurements, reflection-less contacts are essential because any contact reflection would introduce reduction of the transmitted conductance, masking the intrinsic sum rule of compressible QH fluids. The reflection-less criteria is same as for L-B formalism i.e. transmittance from contact to 2DEG and vice versa is always 1. In a reflection-less contact, resistance is expected to be very low compared to the QH resistance (of the order of K$\Omega$), which helps in improving the measurement accuracy of transmitted conductance. Not only that, its temperature variation also is expected to be low for low contact resistance. 

In two-terminal measurement configuration, connecting wire resistances and contact resistances of the 2DEG device introduce systematic drift in the data. For systematic normalization of the conductance data, we consider a resistance $R$ connected to each contact, that is sum of the wire resistance $R_w$ and contact resistance $R_c$. Because of this resistance $R$, the contacts 1, 2 and 3 are not properly grounded and there is a current from contact 1 to 2 etc. Also the voltage at source $S$ is lower than the applied voltage $V_s$ by amount $R\times$(current at source). Considering $G_1 \gg G_2, G_3$, estimated conductance are defined as
\begin{equation}
G_1 \simeq \frac{G_{1m}}{1-2RG_{1m}},
\qquad
G_2 \simeq \frac{G_{2m}}{1-2RG_{1m}},
\qquad
G_3 \simeq \frac{G_{3m}}{1-RG_{1m}},
\end{equation}
where $G_{im}=I_i/V_s$ is the measured conductance at $i^{th}$ contact. Hence, the total measured conductance data $\sum_i G_{im}$ possess systematic deviation at higher conductance value as shown in Fig.~\ref{Fig_S6}. The value of $R$ is found to be nearly $110~\Omega$, calculated from the difference of $(G_{1m}+G_{2m}+G_{3m})$ and $G_H$ for $\nu=3$ plateau for low density sample at 30~mK temperature from Fig.~\ref{Fig_S6}. The same value of $R=110~\Omega$ is used for the estimation of all transmitted conductance values in this letter. Notably transmitted conductance $G_3^{1,2\mathrm{float}}$ is normalized as
\begin{equation}
G_3^{1,2\mathrm{float}}
\simeq
\frac{G_{3m}^{1,2\mathrm {float}}}
{1-3RG_{3m}^{1,2\mathrm{float}}}
\end{equation}
in Fig.~\ref{Fig.2}b of the main text, where
\begin{equation}
G_{3m}^{1,2\mathrm{float}}=
\frac{I_3^{1,2\mathrm{float}}}{V_s}
\end{equation}
is the measured conductance when contacts 1 and 2 are floated.
\begin{figure*}[h]
\includegraphics[width=0.40\textwidth]{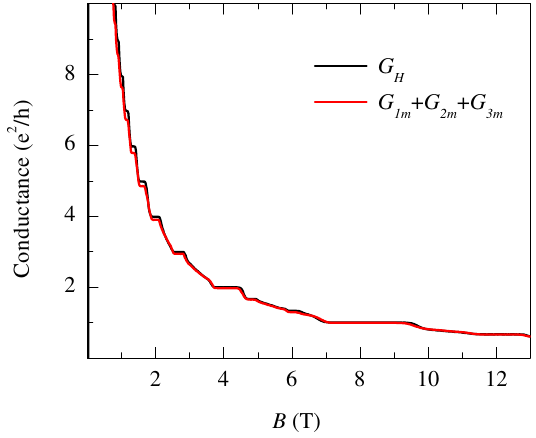}
 \caption{Plot of measured total conductance and Hall conductance data with magnetic field for low density sample at 30 mK.}\label{Fig_S6}
\end{figure*}
In this Hall bar device, contact resistance is about $R_c\sim 10 \Omega$, while measured wire resistance is nearly $R_w\sim 100 \Omega$. Such low contact resistance in a high mobility 2DEG can easily be achieved by standard Au/Ge/Ni contacts made by diffusion.

\begin{center}
	\specialsection{\, Variation of transmitted conductances with magnetic field under dark condition} 
\end{center}
\label{sec: individual G_30 mK_Dark}
Estimated transmitted conductances $G_1$, $G_2$, $G_3$ and total sum of conductance $G_{sum} (= G_1 + G_2 + G_3)$ are plotted with respect to magnetic field sweep from 0 to 13T in Fig.~\ref{Fig_S7} under dark condition at 30 mK temperature. Conductance G1 is dominating and it captures chiral edge conductance as well as a small portion of bulk contribution. Notably, conductance $G_2$ is very small compared to $G_1$ in the QH regimes, where current from contact 1 (contact 1 is not 100$\%$ virtually grounded because of series resistance $R = 110\Omega$) and current from source S through bulk extended states are reaching contact 2 (see Fig.~\ref{Fig.1}b).
\begin{figure*}[h]
\includegraphics[width=0.45\textwidth]{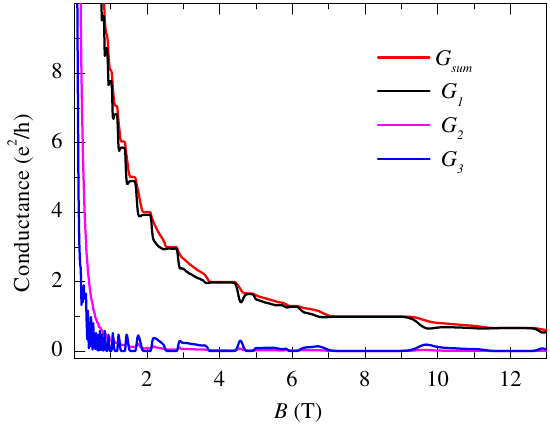}
 \caption{Plot of transmitted conductances with magnetic field at 30 mK temperature under dark condition.}\label{Fig_S7}
\end{figure*}
At the incompressible QH plateaus, since bulk is insulating, no current will pass through the bulk; hence $G_3$ is found always be zero and $G_t$ is quantized in the QH plateaus (Fig.~\ref{Fig.1}c, and Fig.~\ref{Fig_S7}). The result also confirms that there is no edge contribution in conductance $G_3$. Whenever, the system comes out of the QH plateaus, the bulk extended states conduct and conductance $G_3$ becomes finite because of conduction through the partially filled LL. Hence, $G_3$ represents transport of the pure bulk extended states. The value of $G_3$ is smaller than the conductance $G_1$, but it is not at all negligible. Without accounting $G_3$, sum rule of conductance cannot be justified as evident in Fig.~\ref{Fig_S7}. Here, edge/bulk contributions can not be neglected; rather, we demonstrate that in the compressible regime, the edge and bulk contributions are additive and collectively satisfy a universal sum rule, a behaviour that is impossible if the bulk transport is diffusive or percolative.

\begin{center}
	\specialsection{\, Variation of transmitted conductances for reversed magnetic field under dark condition} 
\end{center}
\label{sec: individual G(-B)_30 mK_Dark}
Similar to Fig.~\ref{Fig_S7}, the estimated transmitted conductances are plotted by reversing the magnetic field in Fig.~\ref{Fig_S8} under dark condition. Here large $G_3(-B)$ value indicates change of chirality direction and most of the current arrive at contact 3 due to small travel distance of 20 $\mu$m from the source S and very small amount of currents reach to contacts 1 and 2 in the QH regime. In this measurement condition, Hall resistance could not be measured. Therefore, the total sum of conductances $G_{sum}(-B)$ is compared with the measured Hall conductance $G_H$  under positive magnetic field, where $G_H(+B)$ is obtained from plot in Fig.~\ref{Fig_S4}.
\begin{figure*}[h]
\includegraphics[width=0.40\textwidth]{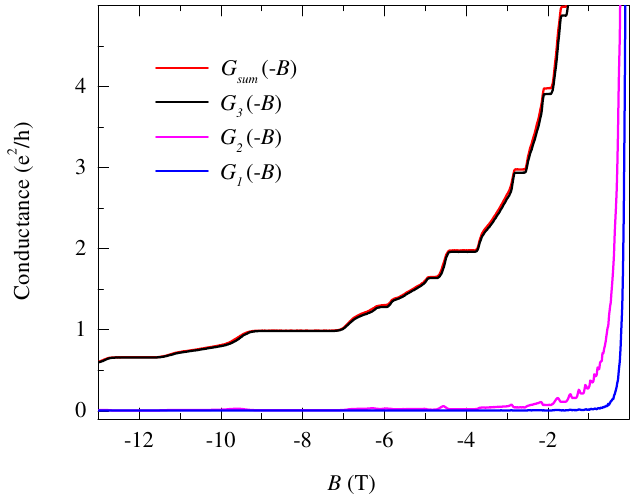}
 \caption{Plot of the transmitted conductances under negative magnetic field in dark condition at 30 mK temperature.}\label{Fig_S8}
\end{figure*}
\newpage
\begin{center}
	\specialsection{\, Presentation of conductances at different temperatures after illumination} 
\end{center}
\label{sec: G_all T_Illumination}
Representative temperature dependence data of transmitted conductances are plotted at temperatures 100, 200, 300 and 400 mK temperature for high density sample in Fig.~\ref{Fig_S9}. At all temperatures, the total sum of transmitted conductances $G_{sum}$ is approaching the Hall conductance $G_H$. The result can not be explained by classical Drude transport, where Hall conductance becomes (by classical definition) $G_H = (nh/eB)(e^2/h) = ne/B$, which does not correspond to the experimental results even at higher temperatures.
\begin{figure*}[h]
\includegraphics[width=0.55\textwidth]{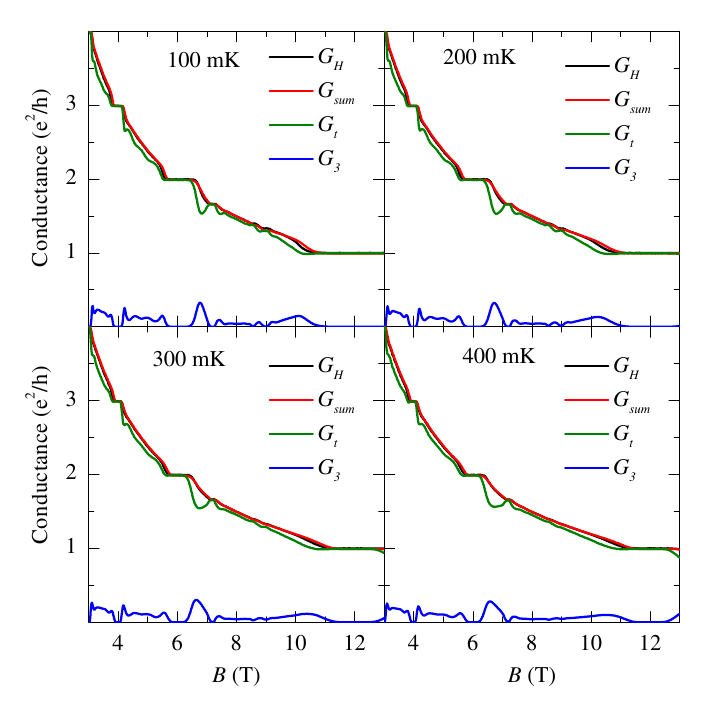}
 \caption{Plot of conductances with magnetic field at 100, 200, 300, 400 mK temperatures after light illumination.}\label{Fig_S9}
\end{figure*}

\begin{center}
	\specialsection{\, Comparison of conductances at different temperatures} 
\end{center}
\label{sec: G_t & G_3_all T_Illumination}
\begin{figure*}[h]
\includegraphics[width=0.55\textwidth]{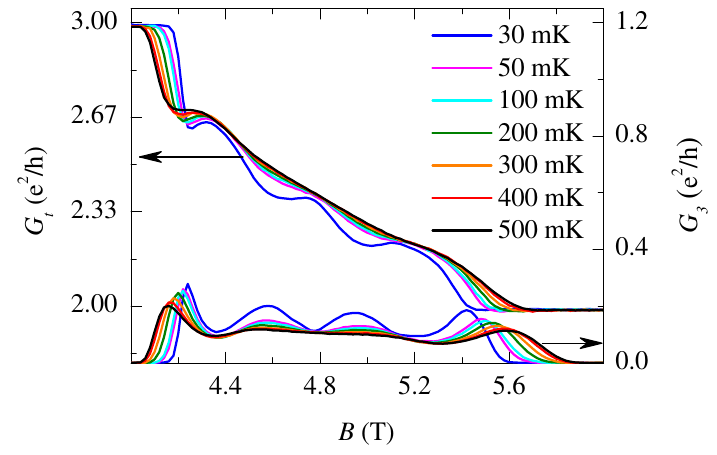}
 \caption{Plot of transmitted conductance $G_t$ and $G_3$ within 4-6~T magnetic field at different temperatures after light illumination.}\label{Fig_S10}
\end{figure*}
Fig.~\ref{Fig_S10} shows the variation of transmitted conductance $G_t$ along the Hall bar and transmitted conductance $G_3$ along the width of the Hall bar at different temperatures within magnetic field range 4-6~T after carrier injection. Because of formation of $8/3$, $5/2$, $7/3$ QH states, deeps in $G_3$ are observed and the deeps are disappearing with increasing temperatures.

\begin{center}
	\specialsection{\, Estimation of Root Mean Square Percentage Error (RMSPE)} 
\end{center}
\label{sec: RMSPE}
To check goodness of overlapping two conductance curves ($G_a$ and $G_b$), we use Root Mean Square Percentage Error (RMSPE) estimated from the data points of the two traces as
\begin{equation}
\mathrm{RMSPE}
=
100\%
\sqrt{
\frac{1}{N}
\sum_{i=1}^{N}
\frac{\left(G_a^{i}-G_b^{i}\right)^2}
{\left(G_b^{i}\right)^2}
}
\end{equation}
where $i$ is the data point number.
\\

\begin{center}
	\specialsection{\, Origin of finite longitudinal resistance under dissipation-less bulk transport} 
\end{center}
\label{sec: origin_R_xx}

Under suppressed dissipative transport within bulk extended states, observing finite longitudinal resistance $R_{xx}$ is counterintuitive in the compressible QH fluids. To clarify the origin of finite longitudinal resistance under dissipation-less transport, we present an example of back reflection of current in a QH circuit at bulk filling fraction $\nu_b=2$ as schematically shown in Fig.~\ref{Fig_S11}. In the circuit, the inner integer mode is back reflected controllably by setting filling fraction $\nu_g=1$ beneath the gates placed across the width of Hall bar. There are six contacts $C1$-$C6$ for the measurements of Hall and longitudinal voltages under excitation voltage $V_S=1$ (dimensionless) applied at the source contact $S$ and drain contact $D$ remains virtually grounded by the current-to-voltage preamplifiers. At each contact, total incoming current is equipartitioned into the two outgoing unity conductance modes. During this process, electro-chemical potential equilibration takes place at the contacts and final electro-chemical potential of the contact is determined by the outgoing current in the modes.

For example, unity conductance ($G=1$) of the modes gives one unit current in each mode at contact $C1$ (for $V_S=1$) of Fig.~\ref{Fig_S11}. At contact $C1$, electro-chemical potential becomes $V_S$ naturally. The inner mode coming out from contact $C1$ is back reflected by the gate, reaching contact $C6$. The outer mode carrying current $i'_1$ is reaching to contact $C6$ of Fig.~\ref{Fig_S11}. After equipartitioning of total incoming current, the outgoing current in each mode becomes
$i_3=\frac{1+i_1'}{2}$.
Hence, electro-chemical potential at the contact $C6$ becomes
$V_{C6}=i_3=\frac{1+i'_1}{2}$.
Similarly at contact $C2$, incoming current $1$ and $i_1'$ is equipartitioned and gives electro-chemical potential at the contact $C2$ as
$V_{C2}=\frac{1+i_1'}{2}=i_1$...(equation 1).
As well, electro-chemical potential at the contact $C5$ becomes
$V_{C5}=i_1'=\frac{i_1}{2}$...(equation 2).
The electro-chemical potential at the contact $C3$ becomes
$V_{C3}=i_2=\frac{i_1}{2}$, while electro-chemical potential at the contact $C4$ becomes $V_{C4}=0$. Solving Eqs.~(1) and (2), we get $i_1'=1/3$ and $i_1=2/3$. Using these solutions, we get electro-chemical potentials of the contacts as follows:
$V_{C1}=V_S, V_{C2}=\frac{2}{3}V_S, V_{C3}=\frac{1}{3}V_S$, $V_{C4}=0, V_{C5}=\frac{1}{3}V_S, V_{C6}=\frac{2}{3}V_S$.
Here longitudinal voltage is coming out to be symmetric as
$V_{C1}-V_{C2}  = V_{C6}-V_{C5}
= \frac{1}{3}V_S$
and $V_{C2}-V_{C3} = V_{C5}-V_{C4} = \frac{1}{3}V_S$.
Notably, the Hall voltage is uniform over the sample:
$V_{C1}-V_{C6} = V_{C2}-V_{C5} = V_{C3}-V_{C4} = \frac{1}{3}V_S$.
In the QH circuit of Fig.~\ref{Fig_S11}, the transmitted current is $2i_3=i_1=2/3$. Therefore, current partitioning is originating finite Hall and longitudinal voltages in the dissipation-less QH circuits because of electro-chemical potential equilibration at the contacts.
\\
\begin{figure*}[h]
\includegraphics[width=0.55\textwidth]{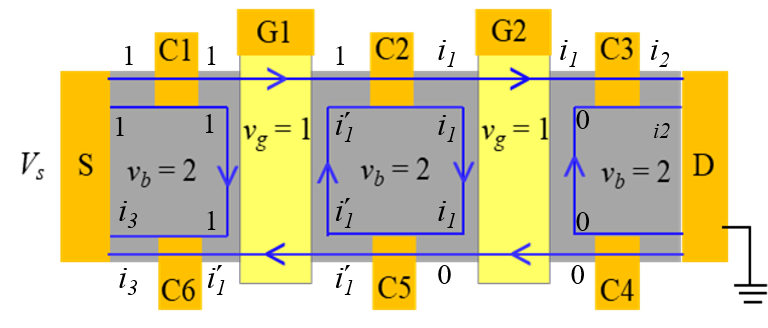}
 \caption{Schematic of Hall circuit with back reflection of inner edge mode at bulk filling fraction $\nu_b=2$. Source $S$ is excited with voltage $V_S=1$ (dimensionless) and drain $D$ is grounded. $C1$-$C6$ are the Ohmic contacts for voltage measurement $V_{C1}$--$V_{C6}$. $G_1$ and $G_2$ are the top gates used for tuning the filling fraction $\nu_g=1$ beneath the gates. The blue lines with arrows indicate current flow through the integer edge modes; where $1$, $i_1$, $i_2$, $i_3$, $i'_1$, and $0$ represent current carried by the edge modes.}\label{Fig_S11}
\end{figure*}
Similarly, in the compressible QH regime, the bulk current and the edge current are partitioned at the contacts, and equilibration of electro-chemical potential at the contacts~\cite{Jain1990_NL_PRL} originates finite longitudinal voltages. Since bulk current is distributed over the sample, hence for a comprehensive understanding of finite longitudinal resistance numerical study is needed utilizing our framework of contact equilibration. However, qualitative understanding of longitudinal resistance $R_{xx}$ under dissipation-less transport due to electro-chemical potential equilibration at the contacts is obtained. From the $R_{xx}$ value, the conductance $G_{xx}$ can easily be calculated. Finally, dimensionless parameter of the system, namely longitudinal conductivity $\sigma_{xx}$, can also be defined. Therefore, the longitudinal conductivity $\sigma_{xx}$ is not arising from dissipation; it is originating from electro-chemical potential equilibration within partially filled LL at the contacts. As a result, $G_{xx}$, $G_3$ and $\sigma_{xx}$ are in the same footing of transport (in chiral metal phase) within partially filled LL under electro-chemical potential equilibration at the contacts. More specifically, current partitioning takes place in the device at the source $S$ and contact arms based on the conductance ratio of the edge modes and the bulk states. The partitioned current in the bulk states mostly reaches to contact 3, resulting in conductance $G_3$. Hence, $G_3$ is also originating, similar to $G_{xx}$, from current partitioning due to electro-chemical potential equilibration at the contacts.

With increasing magnetic field, contribution of bulk conductivity $\sigma_{xx}$ becomes more and more significant (ratio $^{norm}G_{xx}/G_t$ increases at the peak of $G_3$ according to Fig.~\ref{Fig.2}a). Hence, the current loss/gain in the bulk is higher at high magnetic field at the contacts. As a result, peak value of longitudinal resistance $R_{xx}$ is higher at high $B$ because of higher longitudinal voltage $V_{xx}$ under electro-chemical potential equilibration at the contacts (see Fig.~\ref{Fig_S2} and Fig.~\ref{Fig_S3}). This observation is highly consistent with the conjecture of electrochemical potential equilibration at the contacts.

\begin{center}
	\specialsection{\, Broader significance of the study} 
\end{center}
\label{sec: Significance}
We have highlighted the novelty and broader significance as follows:
\\
\textbf{Conduction law of chiral metal phase:} Quantized Hall resistance characterizes the incompressible quantum Hall regime; analogously, our discovered sum rule serves as the defining universal transport signature of the gapless “chiral metal phase”.
\\
\textbf{Resolution of a decades-old transport mystery:} It explains why longitudinal resistance in compressible states doesn't always scale with sample length as expected in a Drude metal. Here, longitudinal resistance does not originate from bulk dissipation but primarily from the equilibration of electrochemical potential at the contacts.
\\
\textbf{Experimental verification of the chiral bulk:} Chiral transport in compressible quantum Hall fluids, characterized by suppressed dissipative transport with a distinct handedness, is confirmed by floating contact measurements. Consequently, this sum rule emerges as a fundamental conduction law of the “chiral metal phase”.
\\
\textbf{Boundary conditions of chiral metal phase:} The conditions of realizing the chiral metal phase in gapless 2D systems are (i) broken time-reversal symmetry at finite magnetic fields and (ii) effective screening ($l_s<l_B$), which transforms the disorder landscape into a flat field with point-like disorder.
\\
\textbf{Robustness across strongly interacting regimes:} The sum rule holds identically across both integer and fractional compressible regimes including highly interacting even denominator states (see Fig.~\ref{Fig.2}c). The result proves that the screening condition ($l_s<l_B$) flattens the disorder landscape effectively enough to enforce unidirectional chiral trajectories under time reversal symmetry breaking even for these strongly interacting composite quasi-particles.
\\
\textbf{A broad paradigm for 2D correlated systems:} Our study introduces a new paradigm of chiral transport across a broad class of gapless two-dimensional systems characterized by short-range screening and broken time-reversal symmetry. By establishing a new paradigm where chirality and disorder compete in gapless states, our work carries immediate, broad implications for researchers studying transport in graphene, moiré superlattices, anomalous Hall materials etc. 
\\
\textbf{Future chiral electronics in semiconductor systems:} At zero magnetic fields, Drude dissipative transport is well established to be isotropic. Hence, our work opens a question about isotropic to chiral metal phase transition with increasing magnetic field. This chiral metal phase might pave a way for future chiral electronics in semiconductor systems.
\\ 
\textbf{Strength of floating contact measurements:} Our findings confirm that the QH fluid is always chiral irrespective of the compressible and incompressible regions at high magnetic fields. This implies that the probability of transmission is $100~\%$ in the crirality direction. Hence, from 1 to 2 to 3 is an easy direction, where $G_3^{1,2float}$ approaches Hall conductance $G_H$ in Fig.\ref{Fig.2}b. While, 3 to 2 to 1 direction is opposite to the chirality. As a result, current from 3 to 2 and 2 to 1 can not reach that is experimentally verified as $G_t^{3float}$ approaches $G_t$ in Fig.\ref{Fig.2}a and $G_1^{2float}$ approaches $G_1$. Most interestingly, the results are valid in the compressible region, and hence the compressible bulk is also chiral.

\end{onecolumngrid}	

\end{document}